\documentclass[prl,aps,reprint]{revtex4-1}

\usepackage[normalem]{ulem}
\usepackage{graphicx} 
\usepackage[usenames]{color}
\usepackage{amsmath,amssymb}
\usepackage{gensymb}
\usepackage{natbib}
\begin{document} 
\title{Why surface nanobubbles live for hours}

\author{Joost H. Weijs}
\affiliation{Physics of Fluids Group, MESA$^{+}$ Institute for Nanotechnology, J. M. Burgers Centre for Fluid Dynamics,
University of Twente, P.O. Box 217, 7500 AE Enschede, The Netherlands}
\author{Detlef Lohse}
\affiliation{Physics of Fluids Group, MESA$^{+}$ Institute for Nanotechnology, J. M. Burgers Centre for Fluid Dynamics,
University of Twente, P.O. Box 217, 7500 AE Enschede, The Netherlands}

\date{\today} 

\begin{abstract}
We present a theoretical model for the experimentally found but counter-intuitive exceptionally long lifetime of surface nanobubbles.
We can explain why, under normal experimental conditions, surface nanobubbles are stable for many hours or even up to days rather than the expected microseconds.
The limited gas diffusion through the water in the far field, the cooperative effect of nanobubble clusters, and the pinned contact line of the nanobubbles lead to the slow dissolution rate.
\end{abstract} 

\maketitle
\paragraph*{Introduction} --
Since their first prediction and discovery almost 20 years ago~\cite{Parker94}, intense research on surface nanobubbles has raised many questions about this intriguing and important phenomenon which has great potential for various applications~\cite{Hampton, Seddonrev11,Craig11,Seddon12_reviewCPC}.
Surface nanobubbles have now been widely reported on various surfaces in contact with water employing various detection mechanisms like Atomic Force Microscopy (AFM) and most recently also through direct optical visualization~\cite{Karpitschka12,OhlPRL12}.
With all these different methods they are found to behave differently than regular macroscopic bubbles.
Surface nanobubbles behave peculiar in several ways: their contact angle is always much lower than expected from Young's law \cite{Borkent, Zhang06ca}, they are stable against violent decompression \cite{BorkentPRL}, and in particular they are stable for much longer than expected: For such small bubbles one would expect a lifetime of order $\mu$s, due to the high Laplace pressure inside the bubbles which drives the gas into the liquid.
On this last question many explanations were proposed, ranging from contamination that shields or limits the diffusive outflux of gas \cite{Ducker09} to a dynamic equilibrium situation where lost gas is replenished \cite{Brenner,SeddonPRL11}.
However, both theories are refuted by experimantal evidence: the addition of surfactants does not influence the behaviour of nanobubbles~\cite{Zhang12_surfactants}, and the circulatory gas flow required for the dynamic equilibrium theory is not measured in all experiments so it cannot be the stabilization mechanism~\cite{SeddonPRL11,OhlPRL12}.
In addition, a large problem with the dynamic equilibrium theory is that it requires some form of driving to satisfy the second law of thermodynamics, and its origin is unclear.
In molecular dynamics, some local inflow near the contact line was indeed observed, but its strength was too weak to explain the stability of surface nanobubbles~\cite{WeijsPRL12}.

A different approach is therefore required, in this Letter we provide an alternative explanation for the long lifetimes of surface nanobubbles.
The theory relies only on classical, well-known, and proven continuum concepts such as diffusion and Henry's law.
Furthermore, the theory only uses confirmed nanobubble properties, namely the pinned contact line \cite{Yang09,ZhangLangmuir12} of nanobubbles and the fact that nanobubbles exist in high coverage fractions at the liquid-solid interface \cite{Yang07,Yang08}.
No fitting or uncontrolled assumptions are required to obtain lifetimes that are consistent with experimental findings.

The Letter is organized as follows. First, the theory is explained and the relevant equations are derived. Next, we solve the equations numerically and analytically.
We vary several parameters to demonstrate the robustness of the long lifetimes of surface nanobubbles in varying experimental conditions.
In addition, we apply the theory to the case of electrolytically generated nanobubbles and find that also here it is consistent with experimental results.
We conclude with predictions from the theory which can straightforwardly be tested in experiments.

\paragraph*{Theory} --
\begin{figure}[t]
\begin{center}
  \includegraphics[width=85mm]{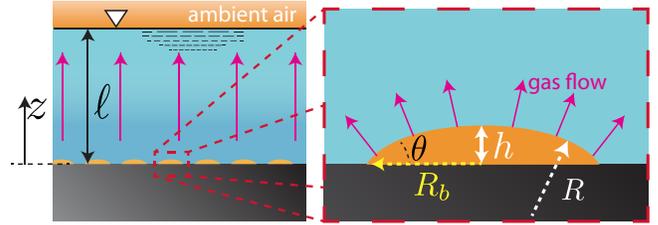}\hspace{3.5cm}
  \caption{(Color online) Sketch of a liquid layer of thickness $\ell$ in contact with a solid (left). The top of the liquid is exposed to atmospheric conditions.  At the solid-liquid interface nanobubbles are present with typical internal contact angle $\theta$, height $h$, and radius of curvature $R$. They are not drawn to scale. The arrows indicate the gas flow direction. On the right a further enlargement of one nanobubble is shown.\label{fig:geometry}}
\end{center}
\end{figure}
We consider an infinitely large plate in contact with a liquid layer with thickness $\ell$ (Fig.~\ref{fig:geometry}).
The liquid layer is in contact with the atmosphere at $z=\ell$ and the solid-liquid interface is located at $z=0$.
The solid-liquid interface is covered with nanobubbles with a number density (per area) of $\rho$, hence the (average) spacing between neighbouring nanobubbles is $\left<d_{bub}\right> \approx \frac{1}{\sqrt{\rho}}$.
In experimental studies nanobubbles are always recovered in high coverage densities~\cite{Tyrrell02, Yang07,Zhang07, Yang08}.
Assuming quasi-steady diffusion, any variation of the dissolved gas concentration $\phi$ in the horizontal ($x,y$) direction due to non-uniform gas outflux decays as $\exp(-\pi z/d_{bub})$.
Hence we can assume that for $\ell > 5\left<d_{bub}\right>$ the the diffusion of gas through the liquid layer is governed by the one-dimensional diffusion equation:
\begin{equation}
\label{eq:diffusion}
\frac{\partial\phi}{\partial t} = D\frac{\partial^2\phi}{\partial z^2}\;,\: z \in [0,\ell]\;,
\end{equation}
with $D$ the diffusion constant of the gas in the ambient liquid and $\phi$ the (number) concentration of gas in the liquid.
The characteristic timescale for diffusion through the water layer is $\ell^2/D\sim10^5$s, which is similar to the lifetimes obtained in experiments.
The vast difference ($\mu$s vs. days) compared to previous estimates originates in using $\ell$ as the relevant length scale instead of the bubble radius $R$, i.e. we use the \emph{far-field} length-scale.
Using $R$ as the length-scale is justified for a (free) bubble in an infinite medium where.
However, in the case of nanobubbles it is important to realize that gas has not left the system until it is released into the atmosphere and hence $\ell$ is the relevant length scale as this is the distance the gas has to travel through the liquid from the bubble towards the atmosphere.
This is also apparant in a thought experiment where we do not allow gas to leave the liquid into the atmosphere, for example by putting the liquid drop in a closed container.
Since the liquid is supersaturated with some gas that has left the bubble, and which cannot escape, an equilibrium is reached (described by Henry's law discussed later in this work) and the bubbles do not dissolve~\cite{WeijsCPC12}.
When opening the container this excess dissolved gas can be released into the ambient air, allowing the bubble to lose gas into the liquid.
This `traffic jam effect' plays a crucial role also in the long nanobubble lifetimes.

We will now fully describe the boundary conditions for above differential equation \eqref{eq:diffusion}. 
They are given by Henry's law, which relates the gas concentration in the liquid to the gas pressure outside the liquid near the interfaces:
\begin{equation}
\label{eq:bc}
\phi(z=0,t)=\frac{p_{bub}(R(N(t)))}{k_H}\;\textrm{~and~}\phi(z=\ell,t)=\frac{p_{atm}}{k_H}\;.
\end{equation}
Here, $k_H$ is Henry's constant, $p_{bub}$ the pressure inside the nanobubbles and $p_{atm}$ the  atmospheric pressure.
$R(t)$ is the radius of curvature of the nanobubbles, which depends on time because the bubbles get flatter as they drain.
In the case of a pinned contact line, the radius of curvature is related to the (internal) contact angle $\theta(t)$ by $R(t)=\frac{R_b}{\sin\theta(t)}$, where $R_b$ is the base radius (cf. Fig.~\ref{fig:geometry}) which is constant due to the pinned contact line.
The (relative) pressure inside the bubbles $p_{bub}(t)$ is the Laplace pressure
\begin{equation}
\label{eq:laplace}
p_{bub}(t)=p_{atm}+\frac{2\gamma}{R}=p_{atm}+\frac{2\gamma}{R_b}\sin\theta(t)\;,
\end{equation}
where $\gamma$ is the liquid-vapour surface tension.
For $\theta<90^\circ$, which is the case for surface nanobubbles, the internal pressure thus decreases as $\theta$ decreases.
This effect provides a negative feedback in the dissolution process, prolonging the lifetime of the nanobubbles.
In this work we do not consider the effects of electrostatic effects on the internal pressure of surface nanobubbles, as electrostatic effects act to reduce the internal pressure and are therefore not a driving force but rather a stabilizing force. It is therefore possible that the derived lifetimes in this work (hours, days) are an underestimation of real lifetimes.

The (single) nanobubble gas content $N(t)$ decreases due to the diffusive flux of gas at $z=0$, which is the location at which gas is transfered from the nanobubbles ($z=0^-$) to the liquid ($z=0^+$).
The diffusive flux is given by Fick's law $J = -D(\partial \phi / \partial z)$, thus
\begin{equation}
\label{eq:flux}
\frac{\textrm{d}N}{\textrm{d} t} =-\frac{J}{\rho}=\frac{D}{\rho}\left.\frac{\partial \phi(z,t)}{\partial z}\right|_{z=0} \;.
\end{equation}
The factor $\rho$ arises to convert the molecular flux per unit area of the substrate to the molecular flux per single surface nanobubble.
Eq.~\eqref{eq:flux} immediately shows how a low nanobubble coverage ($\rho\approx 0$) corresponds to small global flux, $J=-\rho dN/dt$.

To evaluate the boundary condition at $z=0$ (Eq.~\eqref{eq:bc}), we need to relate the number of atoms inside a nanobubble (Eq.~\eqref{eq:flux}) to the geometrical shape of the bubble.
To calculate the geometrical properties of a nanobubble containing $N$ atoms of gas, we use the ideal gas law $p(\theta)V(\theta)=Nk_BT$ using the expression for the volume of a spherical cap $V(\theta) = \frac{\pi}{3}(R_b/\sin\theta)^3\cdot (2-3\cos\theta+\cos^3\theta)$.
Here, $k_B$ is Boltzmann's constant, $T$ the temperature (assumed to be constant at 300K), and $\theta$ the gas-side (internal) contact angle of the nanobubbles.
This implicit equation can be solved numerically for $\theta(N)$.
We can then calculate the radius of curvature of the nanobubbles $R$, which gives us the internal bubble pressure [Eq.~\eqref{eq:laplace}].
Using this pressure, the boundary condition at $z=0$ (Eq.~\eqref{eq:bc}) can be evaluated, which closes our model.
In the next section, we will solve these model equations numerically.

\paragraph*{Numerical evaluation} --
Due to the non-trivial boundary condition at $z=0$ (Eq.~\eqref{eq:bc}) we first solve the diffusion eq.~\eqref{eq:diffusion} numerically.
The simulations were done for different initial conditions.
Since the real initial conditions are unknown, we choose the two extremes between which we expect the real initial conditions.
The first type of initial conditions consists of a linear concentration profile, which allows the system to begin transporting gas from the bubbles immediately ($t=0$),
\begin{equation}
\label{eq:ici}
\phi(z,t=0)=\frac{p_{bub}(R_0)-p_{atm}}{k_H} \left( 1-\frac{z}{\ell}\right)+\frac{p_{atm}}{k_H}\;.
\end{equation}
Here, $R_0$ indicates the initial radius of curvature of the nanobubbles. We choose $R_0$ such that it is equivalent to an initial contact angle $\theta_0=40^\circ$ for given base radius $R_b$.

The second type of initial conditions assumes that the nanobubble formation procedure (ethanol-water exchange or replacing cold water with warm water~\cite{Zhang07}) supersaturates and mixes the water such that the concentration is uniform and equal to the concentration near the nanobubbles:
\begin{equation}
\label{eq:icii}
\phi(x,t=0)=\frac{p_{bub}(R_0)}{k_H}\;.
\end{equation}
The real initial concentration profile will most likely be something between \eqref{eq:ici} and \eqref{eq:icii}: the ethanol-water exchange uniformly supersaturates the water but it takes the nanobubbles some time to form so some gas already drains into the atmosphere.
As we will see, both initial concentration profiles produce long-living nanobubbles.

\begin{figure}[t]
\begin{center}
  \includegraphics[width=88mm]{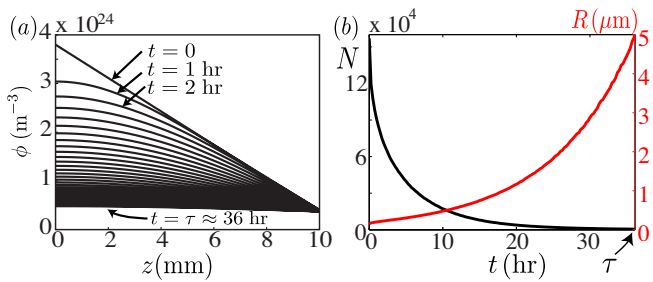}\hspace{3.5cm}
  \caption{(Color online) Results of the calculations using the initial conditions from Eq.~\eqref{eq:ici} and parameters described in the text. (\emph{a}): Snapshots of the concentration profile at 1 hour intervals. Since the bubbles are pinned, the radius of the curvature increases as the bubble drains, lowering the concentration at the bubble side ($z=0$). (\emph{b}): Evolution of the nanobubble gas content (black) and the radius of curvature of the liquid-gas interface $R$ (red) as a function of time.\label{fig:detailedresults}}
\end{center}
\end{figure}
\paragraph*{Results} --
How long does a nanobubble survive according to this description?
Using typical parameters that apply to experiments on surface nanobubbles~\cite{Borkent} ($\tilde\rho=4\cdot 10^{12}$~m$^{-2}$, $\ell=10^{-2}$~m, $\gamma=0.072$~N/m, $D=10^{-9}$~m$^2$/s, $k_H=2.6\cdot 10^{-19}$~Pa$\cdot$m, and $T=300$~K) and the initial condition~\eqref{eq:ici} we obtain the results shown in Fig.~\ref{fig:detailedresults}.
Fig.~\ref{fig:detailedresults}a shows hourly snapshots of the concentration profile $\phi(z)$.
From these curves it is apparent that the transport of gas away from the bubble is limited by the diffusion rate of the gas through water far away from the bubble.
This leads to an almost flat (zero-slope) concentration profile near the bubble through which the diffusive gas flux is very small.
Fig.~\ref{fig:detailedresults}b (red) shows the radius of curvature of the spherical cap $R$ as a function of time.
This radius of curvature increases (due to the pinned contact line) which lowers the internal gas pressure [Eq.~\eqref{eq:laplace}], enhancing the lifetime of the nanobubbles.
Finally, Fig.~\ref{fig:detailedresults}b (black) shows the amount of particles contained inside an individual nanobubble through time.
Here we see that a typical nanobubble which starts out with over 100,000 atoms and ends up with just over 500 atoms after 36 hours.

As a criterion for bubble dissolution we choose a critical bubble height of $h^\ast=1$~nm, after which it may no longer be appropriate to use continuum physics, and also effects such as the disjoining pressure start to dominate the pressure inside the bubble rather than solely the Laplace pressure.
From molecular dynamics it is known that continuum models (e.g. Navier-Stokes, diffusion equation, Henry's law) are valid down to the nanometer scale~\cite{WeijsPRL12,WeijsCPC12}.
Using the same criterion we find that above bubble is stable for 36~hours, which is 10 orders of magnitude longer than previously thought and in agreement with experimental findings and the expected timescale $\ell^2/D$.

\paragraph*{Analytic solution} --
The concentration profiles in Fig.~\ref{fig:detailedresults} suggest that the boundary condition at $z=0$ is of Neumann-type, $\partial\phi/\partial z|_{z=0}=0$.
Taking the time derivative of the boundary condition at $z=0$ [Eq.~\eqref{eq:bc}] and substituting Eq.~\eqref{eq:flux} into the result we obtain
\begin{equation}
\label{eq:bcdirichletornot}
\left.\frac{\partial\phi}{\partial z}\right|_{z=0}=\frac{\rho}{D\frac{d\phi}{dN}}\left.\frac{\partial \phi}{\partial t}\right|_{z=0}\;.
\end{equation}
Filling in representative values for the quantities we find that the concentration gradient at $z=0$ is over 5 orders of magnitude smaller than the typical global concentration gradient ($\approx \hat{\phi}/\ell$), with $\hat{\phi}=\phi-\phi_{atm}$.
This means that the boundary condition for the gradient at $z=0$ can indeed be considered to be approximately zero. For zero gradient the analytic solution is $\hat{\phi}(z,t) = \hat{\phi}_{bub,0}\exp(-\pi^2Dt/(4\ell^2))\cdot\cos(\pi z/(2\ell)) + \hat{\phi}_{transient}(t)$.
The exact form of $\hat{\phi}_{transient}$ depends on the initial conditions, but declines quickly.
A remarkable feature of this result is that $\hat{\phi}(z,t)$ only depends on $D$ and $\ell$, and is completely independent of $\rho$ and $\gamma$.
Of course, 
$\rho$ must be high enough to be able to consider the system as 1-dimensional.
Similarly, the fraction $\rho/(D d\phi/dN)$ must be low enough such that the local gradient at $z=0$ [Eq.~\eqref{eq:bcdirichletornot}] is small compared to the global gradient.

\begin{figure}[th!]
\begin{center}
  \includegraphics[width=88mm]{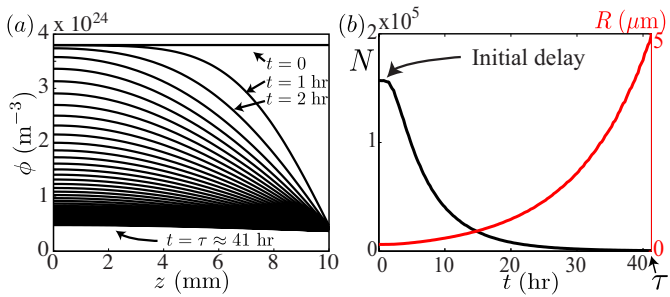}\hspace{3.5cm}
  \caption{(Color online) Numerical results for the initial conditions \eqref{eq:icii}. (\emph{a}): Hourly snapshots of the concentration profile. Due to the initial conditions, some gas that is initially dissolved in the liquid first has to drain to the atmosphere until a concentration gradient exists near the bubble.
As compared to the linear profile case (Fig.~\ref{fig:detailedresults}) the bubbles gain an additional few hours of lifetime because of this. (\emph{b}): Evolution of the nanobubble gas content and the radius of curvature of the surface nanobubbles. 
  \label{fig:altIC}}
\end{center}
\end{figure}

\paragraph*{Varying initial conditions}
How does the initial concentration profile affect the lifetime of the nanobubbles?
During the ethanol-water exchange procedure that is most commonly used to generate nanobubbles experimentally, ethanol is flushed away with clean water.
It is therefore likely that the initial gas concentration profile is uniform in $z$, due to mixing.
We redid the same calculations as before, using an uniform initial concentration profile, the results are plotted in Fig.~\ref{fig:altIC}.
We observe very similar concentration profiles as before, except for small times where the influence of the initial conditions is still felt.
As can be observed in Fig.~\ref{fig:altIC} (b), it takes 2-3 hours before the bubbles `feel' the influence of the ambient air and start to dissolve.
This means that these first hours, the bubbles barely shrink as the (global) concentration gradient near the bubbles is close to zero.

\paragraph*{Robustness of the results} --
How robust are the nanobubble lifetimes against variations in the experimental system?
By changing $\gamma$ and $\ell$ in the numerical calculation, we verify the result from the analytical solution to the diffusion equation for which it holds that the diffusion profile evolution only depends on $D$ and $\ell$.

First, we look at the surface tension $\gamma$. 
Most often an ethanol-water exchange procedure is applied to form nanobubbles.
This method introduces contamination into the system which lowers the surface tension. 
It is therefore important to understand the influence of $\gamma$ on the nanobubble lifetime.
We find that surface tension does not play any role in the dissolution time of nanobubbles.
This result remains counter-intuitive as surface tension is the driving force for nanobubble dissolution.
Indeed, a higher surface tension increases the Laplace pressure [Eq.~\eqref{eq:laplace}], thus increasing the driving that leads to dissolution.
However, it also increases the gas content inside the bubbles.
This denser reservoir requires a larger flux to drain in the same time.
Both effects scale linearly with $\gamma$, hence they cancel.

In previous studies, the liquid height $\ell$ has never been considered as a parameter that affects nanobubble lifetimes.
Based on the analytical solution presented before we expect that $\tau\sim \ell^2/D$, which is indeed exactly recovered from the numerical calculations.
This indicates that the liquid cell size (and geometry) is very important to the lifetime of nanobubbles: the amount of liquid the gas has to travel through determines the timescale of the nanobubble lifetime.
$\ell$ is easy to vary, we suggest to perform corresponding experiments to test our prediction.

\paragraph*{Electrolysis} --
How do nanobubbles behave according to this description when gas is generated at the solid-liquid interface (e.g. by electrolysis~\cite{Zhang06,Yang09})?
A rough estimate based on the values cited in ref.~\onlinecite{Yang09} gives a constant influx of order $\sim 10^5$~molecules per bubble per second.
For an equilibrium to exist (the bubbles neither grow or shrink) the diffusive flux away from the bubble must then be equal to this influx due to electrolysis, hence $\frac{D}{\rho}\hat{\phi}(z=0)/\ell = 10^5$~s$^{-1}$.
This corresponds to a nanobubble contact angle of $\theta=49^\circ$, and nanobubble content $N=2\cdot 10^{5}$.
Interestingly, this means that the bubble contents are refreshed every two seconds.
This only highlights the fact that nanobubbles are not static: without driving (such as electrolysis) they dissolve whereas with driving the gas atoms inside the bubble are replaced every couple of seconds.

\paragraph*{Conclusion} --
We have modelled the gas flow from nanobubbles through the liquid to the atmosphere to study the lifetime of surface nanobubbles.
We find that nanobubbles are not stable, but dissolve by diffusion.
However, due to their pinned contact lines and that the gas has to diffuse towards the atmosphere, they dissolve on a much longer timescale than free bubbles in an infinite liquid.
This last point (diffusion through a relatively thick layer of liquid) explains why previous molecular dynamics results could not recover long nanobubble lifetimes: the system size in molecular dynamics is limited to several tens of nanometers, resulting in lifetimes of order 100~ns, consistent with our findings in ref.~\onlinecite{WeijsPRL12}.
Using the correct length-scales ($\ell\sim 1-10$~mm), we find that surface nanobubbles can easily survive in excess of a day, an increase of 10 orders of magnitudes as compared to previous estimates of $\mu$s.
The results from the experiment where electrolysis is used corroborate the results in this Letter: a non-zero current is measured meaning that gas is continuously formed at the substrate.
We find that the gas flux induced by this current leads to nanobubbles with $\theta=46^\circ$, consistent with experimental results of ref.~\onlinecite{Yang09}.
 \\
J.H. Snoeijer, H. Gelderblom, and X.H. Zhang are gratefully acknowledged for discussions. This work is part of the research programme of the Foundation for Fundamental Research on Matter (FOM), which is part of the Netherlands Organisation for Scientific Research (NWO).


\begin{thebibliography}{10}

\bibitem{Parker94}
J. Parker, P. Claesson, and P. Attard, {\em {Bubbles, cavities, and the
  long-ranged attraction between hydrophobic surfaces}}, {J. Phys. Chem.} {\bf
  {98}},  8468  ({1994}).

\bibitem{Hampton}
M.~A. Hampton and A.~V. Nguyen, {\em Nanobubbles and the nanobubble bridging
  capillary force}, Adv. Colloid Interface Sci. {\bf 154},  30  (2010).

\bibitem{Seddonrev11}
J.~R.~T. Seddon and D. Lohse, {\em {Nanobubbles and micropancakes: gaseous
  domains on immersed substrates}}, {J. Phys.: Condens. Matter} {\bf {23}},
  ({2011}).

\bibitem{Craig11}
V.~S.~J. Craig, {\em Very small bubbles at surfaces-the nanobubble puzzle},
  Soft Matter {\bf 7},  40  (2011).

\bibitem{Seddon12_reviewCPC}
J.~R.~T. Seddon, D. Lohse, W.~A. Ducker, and V.~S.~J. Craig, {\em A
  Deliberation on Nanobubbles at Surfaces and in Bulk}, ChemPhysChem {\bf 13},
  2179  (2012).

\bibitem{Karpitschka12}
S. Karpitschka, E. Dietrich, J.~R.~T. Seddon, H.~J.~W. Zandvliet, D. Lohse, and
  H. Riegler, {\em Nonintrusive Optical Visualization of Surface Nanobubbles},
  Phys. Rev. Lett. {\bf 109},  066102  (2012).

\bibitem{OhlPRL12}
C.~U. Chan and C.-D. Ohl, {\em TIRF Microscopy for the Study of Nanobubble
  Dynamics}, Phys. Rev. Lett.,  in print  (2012).

\bibitem{Borkent}
B.~M. Borkent, S. de~Beer, F. Mugele, and D. Lohse, {\em On the Shape of
  Surface Nanobubbles}, Langmuir {\bf 26},  260  (2010).

\bibitem{Zhang06ca}
X.~H. Zhang, N. Maeda, and V.~S.~J. Craig, {\em Physical Properties of
  Nanobubbles on Hydrophobic Surfaces in Water and Aqueous Solutions}, Langmuir
  {\bf 22},  5025  (2006).

\bibitem{BorkentPRL}
B.~M. Borkent, S.~M. Dammer, H. Sch\"onherr, G.~J. Vancso, and D. Lohse, {\em
  Superstability of Surface Nanobubbles}, Phys. Rev. Lett. {\bf 98},  204502
  (2007).

\bibitem{Ducker09}
W.~A. Ducker, {\em Contact Angle and Stability of Interfacial Nanobubbles},
  Langmuir {\bf 25},  8907  (2009).

\bibitem{Brenner}
M.~P. Brenner and D. Lohse, {\em Dynamic Equilibrium Mechanism for Surface
  Nanobubble Stabilization}, Phys. Rev. Lett. {\bf 101},  214505  (2008).

\bibitem{SeddonPRL11}
J.~R.~T. Seddon, H.~J.~W. Zandvliet, and D. Lohse, {\em {Knudsen Gas Provides
  Nanobubble Stability}}, {Phys. Rev. Lett.} {\bf {107}},    ({2011}).

\bibitem{Zhang12_surfactants}
X. Zhang, M.~H. Uddin, H. Yang, G. Toikka, W. Ducker, and N. Maeda, {\em
  Effects of Surfactants on the Formation and the Stability of Interfacial
  Nanobubbles}, Langmuir {\bf 28},  10471  (2012).

\bibitem{WeijsPRL12}
J.~H. Weijs, J.~H. Snoeijer, and D. Lohse, {\em Formation of Surface
  Nanobubbles and the Universality of Their Contact Angles: A Molecular
  Dynamics Approach}, Phys. Rev. Lett. {\bf 108},    (2012).

\bibitem{Yang09}
S. Yang, P. Tsai, E.~S. Kooij, A. Prosperetti, H.~J.~W. Zandvliet, and D.
  Lohse, {\em Electrolytically Generated Nanobubbles on Highly Orientated
  Pyrolytic Graphite Surfaces}, Langmuir {\bf 25},  1466  (2009).

\bibitem{ZhangLangmuir12}
X. Zhang, D. Chan, D. Wang, and N. Maeda, {\em Stability of Interfacial
  Nanobubbles}, Langmuir,  submitted  (2012).

\bibitem{Yang07}
S. Yang, S.~M. Dammer, N. Bremond, H.~J.~W. Zandvliet, E.~S. Kooij, and D.
  Lohse, {\em Characterization of Nanobubbles on Hydrophobic Surfaces in
  Water}, Langmuir {\bf 23},  7072  (2007).

\bibitem{Yang08}
S. Yang, E.~S. Kooij, B. Poelsema, D. Lohse, and H.~J.~W. Zandvliet, {\em
  Correlation between geometry and nanobubble distribution on HOPG surface},
  Europhys. Lett. {\bf 81},  64006  (2008).

\bibitem{Tyrrell02}
J.~W.~G. Tyrrell and P. Attard, {\em Atomic Force Microscope Images of
  Nanobubbles on a Hydrophobic Surface and Corresponding Force−Separation
  Data}, Langmuir {\bf 18},  160  (2002).

\bibitem{Zhang07}
X.~H. Zhang, X. Zhang, J. Sun, Z. Zhang, G. Li, H. Fang, X. Xiao, X. Zeng, and
  J. Hu, {\em Detection of Novel Gaseous States at the Highly Oriented
  Pyrolytic Graphite−Water Interface}, Langmuir {\bf 23},  1778  (2007).

\bibitem{WeijsCPC12}
J.~H. Weijs, J.~R.~T. Seddon, and D. Lohse, {\em Diffusive Shielding Stabilizes
  Bulk Nanobubble Clusters}, ChemPhysChem {\bf 13},  2197  (2012).

\bibitem{Zhang06}
L. Zhang, Y. Zhang, X. Zhang, Z. Li, G. Shen, M. Ye, C. Fan, H. Fang, and J.
  Hu, {\em Electrochemically Controlled Formation and Growth of Hydrogen
  Nanobubbles}, Langmuir {\bf 22},  8109  (2006).

\end{thebibliography}
\end{document}